\documentclass[12pt]{article}
\usepackage{amsfonts,amsthm,amsmath,amssymb,upgreek,bm}
\usepackage[bookmarks = false, colorlinks = true, linkcolor = darkblue, citecolor = purple]{hyperref}
\usepackage[paper=letterpaper,margin=.85in]{geometry}
\usepackage{graphicx}
\usepackage{units}
\usepackage{float}
\usepackage[export]{adjustbox}
\usepackage{bbold}
\usepackage{xcolor}
\usepackage[shortlabels]{enumitem}
\usepackage{dsfont}
\usepackage[mathscr]{euscript}
\usepackage[normalem]{ulem}

\newcommand{\be}{\begin{equation}}
\newcommand{\ee}{\end{equation}}
\renewcommand{\tilde}{\widetilde}
\renewcommand{\i}{\mathrm{i}}
\renewcommand{\d}{\mathrm{d}}

\numberwithin{equation}{section}

\usepackage{color}
\definecolor{darkblue}{rgb}{0,0,0.6}
\definecolor{purple}{rgb}{0.4,.2,0.7}
\definecolor{darkgreen}{rgb}{0,0.5,0}

\begin{document}
\thispagestyle{empty}

\vspace*{2.5cm}
\begin{center}

{\bf {\LARGE Sphere and disk partition functions \\ \vspace{7pt} in Liouville and in matrix integrals}}

\begin{center}

\vspace{1cm}

 {\bf Raghu Mahajan, Douglas Stanford, and Cynthia Yan}\\
  \bigskip \rm

\bigskip 

Stanford Institute for Theoretical Physics,\\Stanford University, Stanford, CA 94305

\rm
  \end{center}

\vspace{2.5cm}
{\bf Abstract}
\end{center}
\begin{quotation}
\noindent

We compute the sphere and disk partition functions in semiclassical Liouville and analogous quantities in double-scaled matrix integrals. The quantity $\text{sphere}/\text{disk}^2$ is unambiguous and we find a precise numerical match between the Liouville answer and the matrix integral answer. An application is to show that the sphere partition function in JT gravity is infinite.

\end{quotation}

 \newcommand{\psltwor}{\text{PSL}(2,\mathbb{R})}
  \newcommand{\psltwoc}{\text{PSL}(2,\mathbb{C})}
 \newcommand{\volhtwo}{\text{vol}(H^2)}
 \newcommand{\osp}{\text{OSp}(1|2)/\mathbb{Z}_2}

\setcounter{page}{0}
\setcounter{tocdepth}{3}
\setcounter{footnote}{0}
\newpage

\setcounter{page}{2}
\tableofcontents

\section{Introduction}
In string theory, the sphere partition function without operator insertions is a fundamental but confusing quantity. In principle, it should give minus the classical value of the on-shell action of the string background. In the simplest string backgrounds, like empty flat spacetime, this action vanishes. However, there are backgrounds of critical string theory where the on-shell action is nonzero and physically important, like for thermal AdS${}_3\times S^3\times X$. 

It isn't known how to compute the sphere partition function in such cases. Part of the puzzle is that one has to divide by the volume of the conformal Killing group $\psltwoc$, which has infinite volume and no sensible finite regularized value \cite{Liu:1987nz}\cite{Eberhardt:2021ynh}. Another aspect is that in cases where the partition function is expected to be nonzero, the target space is noncompact, and there is a divergent integral over the location of the string worldsheet.

Some proposals exist in the literature. Tseytlin \cite{Tseytlin:1988tv}\cite{Tseytlin:1988rr} has proposed to replace the division by the divergent volume of $\psltwoc$ by a derivative wrt the worldsheet UV cutoff. See also \cite{Callan:1986jb} and \cite{Susskind:1994sm}. Another proposal \cite{Kutasov:1999xu} was that at least for the case of AdS$_3$, the noncompactness and the $\psltwoc$ could cancel each other directly. This proposal was questioned in \cite{Kraus:2002cb}.

It seems likely that the on-shell sphere partition function really is zero up to effects having to do with the noncompactness of the target space. This is consistent with the fact that in the gravity theory one gets from the low-energy limit of string theory, the on-shell action vanishes up to boundary terms \cite{Tseytlin:1988tv}\cite{Gutperle:2002ki}. In that theory, to compute the on-shell action of a noncompact spacetime, one has to put some kind of radial cutoff, add the GHY boundary term together with additional counterterms, and take a limit. Perhaps the resolution of the sphere puzzle from the worldsheet perspective will involve a similar procedure. This seems like a technical challenge: how does one put a spacetime radial cutoff in the worldsheet path integral?

We don't know how to do this. But as a (possibly irrelevant) warmup, in this paper we will discuss an example from noncritical string theory, first studied by Zamolodchikov \cite{Zamolodchikov:1982vx}, where the answer for the sphere partition function is finite and nonzero. Specifically, we study the noncritical string theory consisting of Liouville theory and the $(2,p)$ minimal model, and we do the Liouville path integral directly in the semiclassical limit $c_\text{Liouville} \to +\infty$,  which is relevant for large values of $p$. As shown in \cite{Zamolodchikov:1982vx}, in this example the conformal symmetry of the Liouville theory is spontaneously broken by a semiclassical saddle point, leading to Goldstone zero modes. These zero modes are noncompact, and the integral over them cancels against the divergent volume of $\psltwoc$ that we are supposed to divide by, giving a finite and nonzero answer. So indeed, the noncompactness and the $\psltwoc$ cancel each other neatly.

We do the calculation from \cite{Zamolodchikov:1982vx} in a bit more detail, and match the answer to the predictions of the matrix integral. To do this matching, it is important to compare not the sphere partition function itself, but the well-defined quantity sphere/disk$^2$, where we consider the specific case of the disk with FZZT boundary conditions. So one also has to compute this disk partition function. This is structurally similar to the sphere, with $\psltwor$ playing the role of $\psltwoc$.

In fact, this comparison with the matrix integral was already done (with ZZ disk instead of FZZT) in the work of Alexandrov, Kazakov and Kutasov \cite{Alexandrov:2003nn}, using exact Liouville methods \cite{Dorn:1994xn}\cite{Zamolodchikov:1995aa}\cite{Fateev:2000ik}\cite{Zamolodchikov:2001ah}. Specifically, \cite{Alexandrov:2003nn} gets a formula for the sphere partition function by integrating the the DOZZ formula for the three point function of cosmological constant operators, which represent derivatives with respect to $\mu$ of the sphere partition function.\footnote{This procedure would also work for all the $(q,p)$ minimal string theories and also the $c=1$ string theory.} Our direct semiclassical calculation is a less complete match to the matrix integral, because it is only valid for large $p$. However, it has the advantage that the role of $\psltwoc$ and $\psltwor$ are more obvious. Also, we are able to determine the overall numerical coefficient, which was fitted in \cite{Alexandrov:2003nn}.

On the matrix integral side, defining the analog of the sphere and disk partition functions requires some care, and we explain this in detail. We find that in the matrix integral dual to the $(2,p)$ minimal string, which has the leading density of eigenvalues 
\be
\rho(E) = \frac{e^{S_0}}{2\pi^2}\sinh\left[\frac{p}{2}\text{arccosh}\left(1 + \frac{8\pi^2}{p^2}E\right)\right],
\ee
the universal part of the matrix integral free energy is
\be\label{sphereansintro}
\log(\mathfrak{Z}) \supset -\frac{e^{2S_0}}{2^{10}\pi^6}\frac{p^5}{p^2-4}.
\ee
In the string theory, the density of states is related to the disk with FZZT boundary conditions, and the free energy is the sphere partition function. Our Liouville formulas for these quantities agree with (\ref{sphereansintro}) in the large $p$ limit where one ignores the ``$-4$'' in the denominator. Note that in the strict large $p$ limit, in which this system approaches JT gravity, the sphere partition function diverges, as suggested in \cite{Maldacena:2019cbz}.

While our work was nearing completion, \cite{Anninos:2021ene}\cite{Muhlmann:2021clm} appeared which also study the semiclassical limit of the sphere partition function in Liouville theory.

\section{Liouville computations}
In this section we will compute Liouville path integrals on the sphere, and on the disk (hemisphere) with FZZT boundary conditions, in a semiclassical approximation at large positive Liouville central charge. In this limit, Liouville theory is weakly coupled, and one can compute the path integral by summing over saddle points and including a one-loop determinant. There are several subtleties involved in getting a well-defined answer from these partition functions, and we will start by explaining these.

First, a problem that one runs into is that the one-loop determinants are infinite, due to the existence of noncompact zero modes. The origin of these zero modes is that the saddle point configurations of the Liouville field spontaneously break the conformal symmetry of the theory, leading to a finite number of Goldstone modes. On the sphere, the globally defined conformal symmetry group is $\psltwoc$, and on the disk it is $\psltwor$, and the zero modes parametrize the quotient space $G/H$ where $G = \psltwoc$ or $\psltwor$ and $H = $ PSU(2) or U(1) is the subgroup of $G$ that preserves the saddle point solutions.

One can get a finite and well-defined answer by computing the ratios 
\be\label{divideBy}
\frac{Z_{\text{sphere}}}{\text{vol($\psltwoc$)}}, \hspace{20pt} \text{and} \hspace{20pt} \frac{Z_{\text{disk}}}{\text{vol($\psltwor$)}}.
\ee
Concretely, the division by the infinite volumes in the denominator is accomplished by omitting the zero modes from the one-loop determinants, and then dividing by the volume of the stabilizer subgroup $H$ that leaves the saddle point invariant (note that $H$ has finite volume).

A second subtlety is that the overall normalization of the path integral is ambiguous, due to (i) the conformal anomaly, (ii) the existence of a finite counterterm proportional to the Euler characteristic, and (iii) an arbitrary choice of measure on the group $G$ whose volume we divide by in (\ref{divideBy}). These ambiguities really exist, but they can be made to cancel out in the ratio
\be\label{ratio}
\frac{Z_{\text{sphere}}}{\text{vol($\psltwoc$)}} \cdot\left(\frac{\text{vol($\psltwor$)}}{Z_{\text{disk}}}\right)^2.
\ee
In order to make the ambiguities (i) and (ii) cancel out, we will use the same metric for the two problems, taking the disk to the be the hemisphere. 
We will also use the same cutoff procedure for the computation of the sphere and the disk one loop determinants.

In order to address (iii), which is the ambiguity in the measure on $G$, we need a principle which chooses related measures on $\psltwoc$ and $\psltwor$. 
One might be tempted to use the fact that $\psltwor$ is a subgroup of $\psltwoc$ and (up to a normalization that cancels out in the ratio) there is a preferred metric on $\psltwoc$ that induces a measure on both spaces. In fact, for our purposes, this is actually not the right answer: instead the metric in the two spaces should be multiplied by a further factor of the volume of the sphere or hemisphere that the theory is defined on. This factor introduces some factors of two relative to the naive guess just described. 

This prescription is the correct one for the application of Liouville theory to noncritical string theory. There one is interested in $Z/\text{vol}(G)$ because in string theory, the conformal symmetry is treated as part of the diffeomorphism and Weyl gauge symmetry. From this perspective, the factors of $1/\text{vol}(G)$ arise from zero mode integrals in the path integral over the $bc$ ghosts, and the factor of the volume of the sphere or hemisphere in the $G$ metric described above arises from the normalization of these zero modes on the two spaces.

Having explained these subtleties, let's now give an overview of the computation and set conventions for Liouville theory. We will define the Liouville field $\sigma$ so that the physical metric is
\be
\d s^2 = e^{2\sigma}\hat{\d s^2}.
\ee
In the explicit computations, we will use the sphere or hemisphere as the reference metric
\be
\hat{\d s^2} = \d\theta^2 + \sin^2(\theta)\d\phi^2, \hspace{20pt} \hat{R} = 2, \hspace{20pt} \hat{K}|_{\text{equator}} = 0.
\ee
It is conventional to write the central charge of Liouville theory as $c = 1 + 6(1/b + b)^2$, and to approach the limit of large $c$ by taking $b$ small. Then the Liouville action is
\begin{align}
I &= \frac{1}{b^2}\left\{\frac{1}{4\pi}\int \sqrt{\hat{g}}\left[(\hat\partial\sigma)^2 + \hat{R}\sigma  + 4\pi\mu e^{2\sigma}\right] + \frac{1}{2\pi}\int \sqrt{\hat{h}}\left[\hat{K} + 2\pi \mu_B e^\sigma\right]\right\}\label{firstLine}\\
&\hspace{20pt}+\frac{1}{4\pi}\int \sqrt{\hat{g}}\hat{R}\sigma + \frac{1}{2\pi}\int \sqrt{\hat{h}}\hat{K}\sigma.\label{secondLine}
\end{align}
The parameter $\mu$ is called the cosmological constant, and the parameter $\mu_B$ is called the boundary cosmological constant. Our conventions for these parameters differ by a factor of $b^2$ from the standard ones in the literature.

As written, the terms on the first line are proportional to $b^{-2}$, and the terms on the second line are of order one. We will treat the theory to one loop order in the small $b$ expansion, which means that we want to compute the order $b^{-2}$ and order one terms in $\log(Z)$. This means that we will need to retain the terms on the second line (\ref{secondLine}). However, to the order that we work, these terms can be treated by first-order perturbation theory, simply evaluating them on the classical solution that is obtained from the leading $b^{-2}$ terms. As long as we remember to do this, we only need to take the first line (\ref{firstLine}) into account in determining the classical solutions and one-loop determinants.

The goal is to compute two different partition functions: (1) the partition function on the sphere, and (2) the FZZT partition function on the disk (hemisphere), with $\mu_B$ fixed.
More precisely, as explained above, we will compute the partition functions divided by the volumes of the respective conformal groups. 
For both cases, the answer for small $b$ is
\be
\frac{Z}{\text{vol}(G)} = \sum_{\text{saddles}}e^{-I_{\text{classical}}} \cdot (\text{one-loop det}')\cdot(\text{gauge-fixing factor}).
\ee
Here the one-loop determinant is computed with the zero modes omitted, and the gauge-fixing factor will convert this prescription into a properly normalized division by $\text{vol}(G)$.

In the rest of the computation, we will go through and evaluate each of these three factors for the sphere and for the disk.

\subsection{Classical solutions and action}
\subsubsection{Sphere}
On the sphere, there is a simple family of classical solutions given by constant configurations of $\sigma$. Restricting to such configurations, the equation of motion (obtained by varying the $b^{-2}$ part of the action with respect to $2\sigma$) is
\be\label{sphereSolutions}
1 + 4\pi \mu e^{2\sigma} = 0, \hspace{20pt} \implies \hspace{20pt} 2\sigma = \log(\frac{1}{4\pi\mu}) + \i\pi(1 + 2n).
\ee
We see that there are actually an integer-indexed family of solutions, in which the Liouville field $\sigma$ differs by $2\pi \i n$. Associated to each constant solution is a family of position-dependent solutions with the same action, obtained by acting with $\psltwoc$ on the constant solutions; we will address these later.

One might be surprised by the fact that there are any classical solutions for Liouville theory on a spherical topology, given that the equations of motion impose that the physical metric $e^{2\sigma}\hat{\d s^2}$ should have constant negative curvature, and that no everywhere-negative-curvature metric is possible on a spherical topology. In fact, for the solutions (\ref{sphereSolutions}), the metric $e^{2\sigma}\hat{\d s^2}$ is a round sphere with an overall negative sign in front. Formally, these solutions have negative curvature $R <0$ and count as valid complex solutions to the equations of motion (this point was explained in JT gravity in \cite{Maldacena:2019cbz}).  
Plugging them into the action, we find
\be
e^{-I_{\text{classical}}} = -e^{-\frac{\i\pi}{b^2}(1+2n)}e^{\frac{1}{b^2}}(4\pi\mu)^{\frac{1}{b^2}+1}.
\ee

Which, if any, of these solutions are we supposed to sum over? If the theory is defined by analytic continuation in $b$, starting from the region where $b$ has a positive imaginary part, then the correct answer is to sum the solutions with $n = 0,1,2,\dots$ \cite{Harlow:2011ny}. This gives the result
\be\label{sumovern}
\sum_{n = 0}^\infty e^{-I_{\text{classical}}} = \frac{\i}{2\sin(\frac{\pi}{b^2})}e^{\frac{1}{b^2}}(4\pi\mu)^{\frac{1}{b^2}+1}.
\ee
In principle, one should sum over the saddle points at the end, after including the one-loop determinants and gauge fixing factors, but these are independent of $n$, so it is allowable to sum over the saddles at this early stage.

To motivate this prescription, one can consider a toy integral
\be
\int \d \sigma e^{a \sigma - e^{2\sigma}}.
\label{toyint}
\ee
After setting $a = -2(b^{-2} + 1)$ and shifting $\sigma$ by a constant, this corresponds to the truncation of the Liouville path integral to the constant mode of $\sigma$. If the real part of $a$ is positive, then the integral converges along the real axis. For our problem, $a$ is negative, and the integral does not converge, but we can imagine defining it by analytic continuation, starting from positive values of $a$, and gradually adjusting the defining contour as we vary $a$ in order to make the integral remain convergent. If we vary $a$ through the upper half plane from positive values almost all the way to the negative real axis, then one acceptable defining contour is the one shown below:
\be\label{contourDiagram}
\includegraphics[width = .5\textwidth, valign = c]{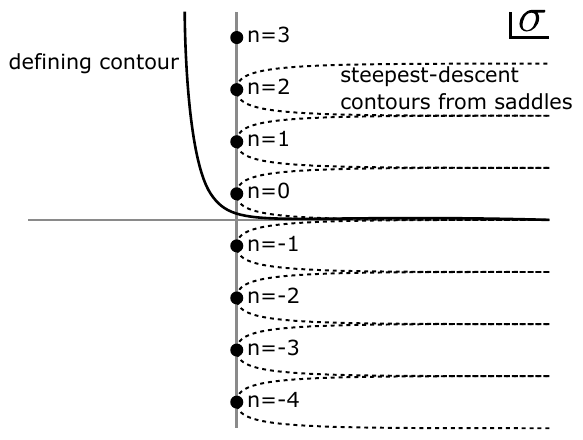}
\ee
Also shown on the diagram are the locations of the saddle points and the steepest-descent contours for each of the saddles (the real part of the locations of these saddles depends on $\mu$). The defining contour is equivalent to a sum of all of the steepest-descent contours for $n = 0,1,2,\dots$, justifying the sum in (\ref{sumovern}).

\subsubsection{FZZT disk}
On the disk (hemisphere), a constant $\sigma$ is not a solution to the equations of motion even with a complex value of $\sigma$. The next simplest thing is to find solutions $\sigma(\theta)$ that are independent of the angular coordinate $\phi$. The Liouville action for such a configuration is
\begin{align}
I &= \frac{1}{b^2}\left\{\frac{1}{2}\int_0^{\frac{\pi}{2}}\d\theta\sin(\theta)\left[\sigma'^2 + 2\sigma  + 4\pi\mu e^{2\sigma}\right] + 2\pi \mu_B e^{\sigma(\pi/2)}\right\}+\frac{1}{2}\int_0^{\frac{\pi}{2}}\d\theta\sin(\theta)2\sigma.
\end{align}
The equations of motion are obtained by varying the first term, of order $b^{-2}$, with respect to $\sigma$. We find the equations
\be
1 + 4\pi \mu e^{2\sigma} = \frac{1}{\sin(\theta)}(\sigma' \sin(\theta))', \hspace{20pt} 2\pi\mu_B e^{\sigma(\pi/2)} + \sigma'(\pi/2) = 0.
\ee
One can check that the following is a solution
\begin{align}
\label{hemisphereSol}
\sigma &= 2\pi \i n-\frac{1}{2}\log(4\pi\mu) + \log\left[\frac{2\alpha}{(1+\alpha^2)\cos(\theta) + (1-\alpha^2)}\right], \\
1 &+ \alpha^2 + 2\sqrt{\frac{\pi}{\mu}}\, \mu_B \alpha = 0. \label{alphamub}
\end{align}
This solution corresponds to the metric $e^{2\sigma}\hat{\d s^2}$ being a piece of the hyperbolic disk, written in a way that is conformal to the hemisphere. The action is given by plugging in and integrating:
\begin{align}
I &= \frac{1}{b^2}\left[2\pi \i n -\frac{1}{2}+\log(\alpha) - \frac{1}{2}\log(4\pi \mu)\right] + \left[-\frac{1}{2}\log(4\pi \mu) +  1 + \frac{2\log\alpha - (1-\alpha^2)\log\frac{2\alpha}{1-\alpha^2}}{1+\alpha^2}\right].
\end{align}

In this expression, $\alpha$ is a parameter of the solution, and is determined by the boundary cosmological constant $\mu_B$ in (\ref{alphamub}). When we compute the one-loop determinants later, we will actually only do the computation in the limit of small positive $\alpha$, which corresponds to large negative $\mu_B$. Physically, this is a high energy limit in the boundary theory (matrix integral). In this limit, we have the leading behavior
\be
e^{-I_{\text{classical}}} = \frac{2}{e}e^{-2\pi \i n/b^2} e^{\frac{1}{2b^2}}\left(\frac{\sqrt{4\pi\mu}}{\alpha}\right)^{\frac{1}{b^2}+1}.
\ee

Again, there is an integer-indexed family of solutions, and one has to decide which solutions should be included. We will assume that it is correct to imitate the case of the sphere, and sum over $n = 0,1,2,\dots$, which leads to the answer
\be
\sum_{n = 0}^\infty e^{-I_{\text{classical}}} = -\frac{\i e^{\i \pi/b^2}}{2\sin(\frac{\pi}{b^2})}\cdot \frac{2}{e}e^{\frac{1}{2b^2}}\left(\frac{\sqrt{4\pi\mu}}{\alpha}\right)^{\frac{1}{b^2}+1}.
\ee
Our understanding of the contour is not as good for this case as for the sphere, but one piece of evidence for this formula is that exact Liouville formulas \cite{Fateev:2000ik}\cite{Kutasov:2004fg} do contain a factor of $1/\sin(\pi/b^2)$, which arises in this expression from the sum over saddles.


\subsection{One loop determinants}
To compute the one-loop determinant, we expand around a classical solution
\be
\sigma = \sigma_{\text{cl}} + \chi
\ee
and integrate over the fluctuation $\chi$ with an appropriate action and measure. The measure is derived from an ultralocal metric in field space
\be\label{metric}
\d s^2 = C^2\cdot (\d\chi,\d\chi) 
\ee
where we introduced an arbitrary constant $C$ to parametrize the normalization ambiguity in the metric, and we defined
\be\label{innerProd}
(f,g) = \frac{1}{4\pi}\int \d^2x\, \sqrt{\hat{g}}\, f(x)g(x).
\ee
The action for the fluctuations is just the quadratic approximation to the full action near the saddle point. This can be written
\be
I \supset \frac{1}{b^2}(\chi,\mathcal{O}\,\chi)
\ee
for a particular differential operator $\mathcal{O}$ that depends on the solution we are expanding around. 

To compute the integral, it is convenient to work in a basis of eigenfunctions of the quadratic action, $\chi(x) = \sum_i \chi_i Y_i(x)$ where
\be
[\mathcal{O} Y_i](x) = \lambda_i Y_i(x), \hspace{20pt} (Y_i,Y_j) = \delta_{ij}.
\ee
In this basis, the field-space metric (\ref{metric}) is
\be
\d s^2 = C^2 \sum_i \d \chi_i^2
\ee
and the path integral is formally
\be\label{oneLoopFormal}
\int \mathcal{D}\chi e^{-\frac{1}{b^2}(\chi,\mathcal{O}\,\chi)} = \prod_i C\int \d \chi_i e^{-\lambda_i \chi_i^2/b^2} = \prod_i \frac{\sqrt{\pi}\,bC}{\sqrt{\lambda_i}}.
\ee

\subsubsection{Sphere}

Expanding around any of the classical solutions (\ref{sphereSolutions}), the quadratic part of the action (\ref{firstLine}) is
\be
I \supset \frac{1}{4\pi b^2}\int \sqrt{\hat{g}}\left[(\partial\chi)^2 -2\chi^2\right].
\ee
The eigenfunctions of this problem are the spherical harmonics, and the eigenvalues are
\be
\lambda = \ell(\ell+1) - 2, \hspace{20pt} \text{degeneracy $2\ell+1$}.
\ee
We have to deal separately with the $\ell = 0$ eigenfunction, the $\ell = 1$ eigenfunctions, and all of the others.

First, note that the $\ell = 0$ eigenfunction is a negative mode, with eigenvalue $\lambda = -2$. This is to be expected based on the diagram (\ref{contourDiagram}). In that diagram, the steepest-descent contours pass vertically through the saddle points, which means that the action is unstable with respect to real perturbations in the constant mode of $\chi$. In the quadratic approximation, the steepest descent contour is just the imaginary axis, and the integral is
\be
C\int_{+\i \infty}^{-\i \infty} \d \chi_0 e^{2 \chi_0^2/b^2} = -\i \frac{\sqrt{\pi}\,bC}{\sqrt{2}}.
\ee

Next, the $\ell = 1$ modes are the zero modes that we promised. These correspond to the Goldstone modes of the $\psltwoc$ symmetry that is spontaneously broken by the constant solutions (\ref{sphereSolutions}). We will take these properly into account in the gauge-fixing part of the computation; for now we simply insert delta functions, so the contribution of these modes is
\be
\prod_{i = 1}^3 C\int \d \chi_i \delta(\chi_i) = C^3.
\ee

Finally, we have the product over all of the other modes with $\ell \ge 2$:
\be
\prod_{\ell = 2}^\infty \left[ \frac{\sqrt{\pi}b C}{\sqrt{\ell(\ell+1)-2}}\right]^{2\ell+1}.
\ee
This is a divergent product, but we can compute a regularized version of its logarithm using the following sums:
\begin{align}\label{sum111}
\sum_{\ell = 0}^\infty (2\ell+1) e^{-\epsilon^2\ell(\ell+1)} &= \frac{1}{\epsilon^2} + \frac{1}{3} + O(\epsilon^2)\\
\sum_{\ell=2}^\infty(2\ell+1)\log\Big[\ell(\ell+1)-2\Big]e^{-\epsilon^2 \ell(\ell+1)} &= \frac{\log(\frac{1}{\epsilon^2}) - \gamma}{\epsilon^2}-2\log(\frac{1}{\epsilon^2}) + 2.32713 + O(\epsilon^2),\label{sum222}
\end{align}
where $\gamma$ is the Euler-Mascheroni constant. We determined the divergent terms by approximating the sums as integrals, and we determined the constant terms numerically.
Using these formulas, one can compute 
\begin{align}
\sum_{\lambda > 0} \log\left[\frac{\sqrt{\pi}b C}{\sqrt{\lambda}}\right]e^{-\epsilon^2 \lambda} &= \sum_{\ell  = 2}^\infty (2\ell+1)\log\left[\frac{\sqrt{\pi}b C}{\sqrt{\ell(\ell+1)-2}}\right]e^{-\epsilon^2 \ell(\ell+1)}\\
&= \log(\sqrt{\pi}bC)\times (\ref{sum111}) - \frac{1}{2}\times (\ref{sum222}).\label{sphereSum}
\end{align}

\subsubsection{Hemisphere}
Expanding around the solution (\ref{hemisphereSol}) to quadratic order, one finds
\begin{align}
I &\supset \frac{1}{b^2}\left\{\frac{1}{4\pi}\int_{\text{hemisphere}}\hspace{-20pt}\sqrt{\hat{g}}\left[(\partial \chi)^2 + \frac{8\alpha^2\chi^2}{((1{+}\alpha^2)\cos(\theta) + (1{-}\alpha^2))^2}\right] - \frac{1}{4\pi}\frac{1+\alpha^2}{1-\alpha^2}\int_{\text{equator}}\hspace{-13pt}\chi^2\right\}\\
&\approx \frac{1}{b^2}\left\{\frac{1}{4\pi}\int_{\text{hemisphere}}\hspace{-20pt}\sqrt{\hat{g}}(\partial \chi)^2  - \frac{1}{4\pi}\int_{\text{equator}}\hspace{-13pt}\chi^2\right\}.
\end{align}
In the second line, we gave an approximate formula for small $\alpha$. The eigenfunctions are determined by solving
\be
-\partial^2\chi = \lambda \chi,
\ee
where $\partial^2$ is the Laplacian on the sphere, and by imposing the boundary condition at the equator
\be
\chi'(\pi/2) = \chi(\pi/2).
\ee
The solutions are
\be
P^{m}_\ell(\cos(\theta))e^{\i m\phi}, \hspace{20pt} \lambda = \ell(\ell+1)
\ee
where $P$ is the generalized Legendre function and $\ell$ is a (non-integer!) parameter that is determined by solving the boundary condition equation.

The spectrum of this operator is qualitatively similar to the one-loop spectrum on the sphere. There is one negative eigenvalue, there are two zero modes, and there are an infinite number of other eigenvalues for which the product requires regularization. The negative mode is in the $m = 0$ sector, and numerically, its eigenvalue is $\lambda_0\approx -1.51095$. Its contribution to the one-loop determinant is
\be
C\int_{+\i \infty}^{-\i \infty} \d \chi_0 e^{-\lambda_0 \chi_0^2/b^2} \approx -\i \frac{\sqrt{\pi}\,bC}{\sqrt{1.51095}}.
\ee
The two zero modes are the lowest eigenvalues in the $m = \pm 1$ sectors, with $\ell = 0$. These correspond to the two spontaneously broken generators of $\psltwor$. 
We will treat these in the gauge-fixing step, but for now we insert delta functions, so they contribute
\be
\prod_{i = 1}^2 C\int \d \chi_i \delta(\chi_i) = C^2.
\ee

Next we discuss the product over all of the other modes. A regularized version of the product (\ref{oneLoopFormal}) can be computed using the following sums
\begin{align}\label{sum1}
\sum_{\lambda>0}e^{-\epsilon^2\lambda} &= \frac{1}{2\epsilon^2} + \frac{\sqrt{\pi}}{4\epsilon} - \frac{11}{6} + O(\epsilon)\\
\sum_{\lambda>0}\log(\lambda) e^{-\epsilon^2\lambda} &= \frac{\log\frac{1}{\epsilon^2} - \gamma}{2\epsilon^2} + \sqrt{\pi}\frac{\log\frac{1}{\epsilon^2} - \log(4) - \gamma}{4\epsilon} +0.57136 + O(\epsilon).\label{sum2}
\end{align}
Here, the sum runs over the positive eigenvalues, omitting the two zero eigenvalues and the one negative eigenvalue. The divergent terms are the same as with Neumann boundary conditions, where the eigenfunctions are a subset of the ordinary spherical harmonics with integer $\ell$. This allowed us to compute the divergent terms analytically by approximating them as integrals. The finite terms in the sums were computed numerically. To get good precision, it was necessary to use the package \url{https://github.com/JamesCBremerJr/ALegendreEval} \cite{bremer2018algorithm} to compute generalized Legendre functions with large parameters.

These formulas can be used to compute the regularized log-determinant
\be
\sum_{\lambda > 0} \log\left[\frac{\sqrt{\pi}b C}{\sqrt{\lambda}}\right]e^{-\epsilon^2 \lambda}.\label{hemisphereSum}
\ee
But after doing so, one finds a problem. For the ratio $Z_{\text{sphere}} / Z_{\text{disk}}^2$, we need the log determinant for the sphere (\ref{sphereSum}) minus two times the log determinant for the disk, (\ref{hemisphereSum}). In this combination, our formulas imply that the leading quadratic divergence cancels out (they can also be individually absorbed using an area counterterm), but the linear and the log divergences remain. The linear divergence can be absorbed into a boundary length counterterm for the disk, but the mismatch in the log term would mean that the ratio is not well-defined.

This problem can be avoided if we regularize the disk in a different way, by inserting in the sum a slightly different convergence factor $e^{-\epsilon^2\tilde{\lambda}_i}$ where
\begin{align}
\tilde{\lambda}_i &= \frac{1}{4\pi}\int \sqrt{\hat{g}} (\partial Y_i)^2\\
&= \lambda_i + \frac{1}{4\pi}\int_{\text{bdy}}Y_i^2.\label{bdyTERM}
\end{align}
One can now recompute the sums (\ref{sum1}) and (\ref{sum2}) using the new regulator. In fact, this isn't much work: the difference $\tilde{\lambda} - \lambda$ is of order one, even for very large eigenvalues, so the change makes a multiplicative correction of order $\epsilon^2$. The only way a small correction like this can affect the order-one terms in the answer is by correcting the leading quadratic divergences. These are dominated by large eigenvalues, so it is enough to know that on average the boundary term in (\ref{bdyTERM}) approaches two for large eigenvalues. (This can be shown by using the fact that for large eigenvalues, the eigenfunctions approach those of the hemisphere with Neumann boundary conditions.)

The upshot is that one finds the revised answers for the sums
\begin{align}\label{sum3}
\sum_{\lambda>0}e^{-\epsilon^2\tilde{\lambda}} &= \frac{1}{2\epsilon^2} + \frac{\sqrt{\pi}}{4\epsilon} - \frac{17}{6}+ O(\epsilon)\\
\sum_{\lambda>0}\log(\lambda) e^{-\epsilon^2\tilde{\lambda}} &= \frac{\log\frac{1}{\epsilon^2} - \gamma}{2\epsilon^2} + \sqrt{\pi}\frac{\log\frac{1}{\epsilon^2} - \log(4) - \gamma}{4\epsilon} - \log(\frac{1}{\epsilon^2})+1.14858 + O(\epsilon).\notag
\end{align}
Now when we subtract twice the disk answer from the sphere answer, the log term cancels, and we can proceed to analyze the finite parts. Note that a priori, both $\lambda$ and $\tilde{\lambda}$ seem like reasonable quantities to use in the regularization, and to  be honest, we would not have known which was right. However, cancellation of the log term seems to require using $\tilde{\lambda}$, and once this choice is made, the finite parts are determined.

\subsection{Dividing by the volume of the conformal group}
Liouville theory on the sphere has an exact $\psltwoc$ conformal symmetry. 
A PSU(2) subgroup of this corresponds to ordinary rotations of the sphere; these symmetries are preserved by the constant saddle points (\ref{sphereSolutions}). However, as we will see, the remaining three directions in $\psltwoc$ are spontaneously broken, which means that if we act with an $\psltwoc$ generator in this subspace, it changes the saddle point nontrivially to a new saddle point with shifted values of the zero modes $\chi_1,\chi_2,\chi_3$ that we found in the one-loop determinant.

We will use coordinates $s_1,s_2,s_3$ for the stabilizer subgroup that is preserved, and $b_1,b_2,b_3$ for the directions that are broken by the classical solution. The $\chi_1,\chi_2,\chi_3$ zero modes can be considered functions of the $b_j$ coordinates.

One can define $Z_{\text{sphere}}/\text{vol}(\psltwoc)$ using the Fadeev-Popov procedure. Starting with the measure for the field zero modes, divided by the measure on $\psltwoc$, we replace it as follows
\begin{align}\label{sphereFP}
\frac{\d(\text{sphere zero modes})}{\d(\psltwoc)} &= \frac{\d \chi_1\d \chi_2\d \chi_3}{\underbrace{\d s_1 \d s_2 \d s_3}_{\text{PSU}(2)}\underbrace{\d b_1\d b_2\d b_3}_{\psltwoc/\text{PSU}(2)} } = \frac{1}{\d s_1\d s_2\d s_3} \text{det}(\frac{\partial \chi_i}{\partial b_j}).
\end{align}
In the final expression, we have an inverse measure on PSU(2) and a Fadeev-Popov determinant. The integral gives the determinant divided by the volume of PSU(2). 

The situation for the disk is very similar to that of the sphere, except that we only have a $\psltwor$ subgroup of the conformal symmetry. The analogous Fadeev-Popov procedure is 
\begin{align}\label{diskFP}
\frac{\d(\text{disk zero modes})}{\d(\psltwor)} &=\frac{ 2^{3/2}\d \chi_1\d \chi_2}{\underbrace{\d s_1}_{\text{U}(1)}\underbrace{\d b_1\d b_2}_{\psltwor/\text{U}(1)} } =  \frac{2^{3/2}}{\d s_1} \text{det}(\frac{\partial \chi_i}{\partial b_j}).
\end{align}
In this case, we end up with a Fadeev-Popov determinant and an inverse measure on U(1). Note that we inserted an important factor of $2^{3/2}$ in this expression relative to (\ref{sphereFP}). This factor will be explained below.

Let's now work out the details explicitly. We write the reference sphere or hemisphere in stereographic coordinates
\be
\hat{\d s^2} = \frac{4 \d z \d \bar z}{(1 + z\bar{z})^2}.
\ee
The infinitesimal $\psltwoc$ or $\psltwor$ transformations correspond to the following set of six holomorphic vector fields ($c$ ghost zero modes)
\be\label{transf}
\begin{tabular}{c|c|c|c}
$a$ & $C_{0,a}^z = \delta_a z$ & $C_{0,a}^{\bar{z}} = \delta_a \bar{z}$ & \text{coordinate} \\
\hline
1 & $\i z$ & $-\i \bar{z}$ & $s_1$\\
2 & $\tfrac{1}{2}(1-z^2)$ & $\tfrac{1}{2}(1-\bar{z}^2)$ & $b_1$\\
3 & $\tfrac{\i}{2}(1+z^2)$ & $-\tfrac{\i}{2}(1+\bar{z}^2)$ & $b_2$\\
4 & $z$ & $\bar{z}$ & $b_3$\\
5 & $\tfrac{\i}{2} (1-z^2)$ & $-\tfrac{\i}{2}(1-\bar{z}^2)$ & $s_2$\\
6 & $\tfrac{1}{2}(1+z^2)$ & $\tfrac{1}{2}(1+\bar{z}^2)$ & $s_3$
\end{tabular}
\ee
The first three of these vector fields preserve the hemisphere $|z|\le 1$, and these correspond to the $\psltwor$ subgroup of $\psltwoc$. 
The last three make sense only on the full sphere. In the final column, we have anticipated results below and labeled the transformations according to whether they are preserved ($s$) or broken ($b$) by the classical solution.

One way to get the right meausure in the $s_i$ and $b_i$ coordinates is to use the perspective of the $bc$ ghost path integral. The determinants of the nonzero modes cancel between the sphere and the disk$^2$ (see e.g.~\cite{Douglas:1986eu}) and the zero modes are integrated with a measure that is given by the square root of the determinant of the field-space inner product of the $c$-ghost zero modes:
\be
M_{ab} = \frac{3}{8\pi}\int \sqrt{\hat{g}} C_{0,a}^\alpha C_{0,b}^{\beta}\hat{g}_{\alpha\beta}.
\label{Mabmatrix}
\ee
The constant $3/8\pi$ out front will cancel out in the ratio sphere$/($disk$)^2$, and we chose it to so that the answer is simply that $M$ is the $6\times 6$ identity matrix for the sphere, and one-half of the $3\times 3$ identity matrix for the disk. So the measure is one for the sphere, and $2^{-3/2}$ for the disk, justifying the numerical factor in (\ref{diskFP}).

It remains to compute the determinants of $\partial \chi_i/\partial q_j$. Conformal transformations are defined to act on the Liouville field in such a way that the physical metric remains invariant. So, under a general 
\be
z\rightarrow \tilde{z}(z)
\ee
we require that
\be
e^{2\tilde{\sigma}(\tilde{z},\bar{\tilde{z}})}\frac{\d \tilde{z} \d \bar{\tilde{z}}}{(1+\tilde{z}\bar{\tilde{z}})^2} = e^{2\sigma(z,\bar{z})}\frac{\d z\d \bar{z}}{(1+z\bar{z})^2}. 
\ee
Infinitesimally, for $\tilde{z}(z) = z + \delta z$, this implies that $\delta \sigma (z) = \tilde{\sigma}(z) - \sigma(z)$ is given by
\be
\delta\sigma = -\left[\delta z \left(\partial_z\sigma - \frac{\bar z}{1+z\bar{z}}\right) + \delta\bar{z} \left(\partial_{\bar{z}}\sigma - \frac{z}{1 + z\bar{z}}\right) + \frac{1}{2}\left(\partial_z\delta z + \partial_{\bar{z}}\delta\bar{z}\right)\right].
\ee
We see that the transformation depends on the classical solution $\sigma$ that we start with. The classical solutions for the sphere and the disk are 
\begin{align}
\text{sphere:} \hspace{20pt} \sigma &= \text{const.}\\
\text{disk:} \hspace{20pt} \sigma &= \text{const.} + \log\frac{\alpha(1+z\bar{z})}{1-\alpha^2z\bar{z}}\approx \text{const.}' + \log(1+z\bar{z})
\end{align}
where we gave the small $\alpha$ limit in the final expression. Plugging in, one finds that for the transformations (\ref{transf}), the corresponding perturbations to $\sigma$ are
\begin{align}
\text{sphere:} \hspace{20pt} \delta_a\sigma &= \left\{0,\frac{z+\bar{z}}{1+z\bar{z}},-\i\frac{z-\bar{z}}{1+z\bar{z}},\frac{-1+z\bar{z}}{1+z\bar{z}},0,0\right\}\\
\text{disk:} \hspace{20pt} \delta_a\sigma &= \frac{1}{2}\left\{0,z+\bar{z},-\i(z-\bar{z})\right\}.
\end{align}
We see that the $a = 1,5,6$ directions are the PSU(2) symmetry directions that stabilize the classical solution, justifying the labeling in (\ref{transf}).

The nonzero $\delta_a\sigma$ functions correspond precisely to the zero modes of the one-loop determinants, but with an arbitrary normalization. The $\chi_i$ coordinates are the coefficients of {\it normalized} zero modes. To see the discrepancy, we can evaluate the matrix $m_{ab}=(\delta_a \sigma,\delta_b\sigma)$  where the inner product is defined in (\ref{innerProd}). One finds
\begin{align}
\text{sphere:} \hspace{20pt} m  &= \frac{1}{3}\text{diag}(0,1,1,1,0,0)\\
\text{disk:} \hspace{20pt} m &= \frac{\log(4)-1}{4}\,\text{diag}(0,1,1).
\end{align}
The determinants $\text{det}(\partial \chi_i/\partial q_j)$ are just the square roots of the determinants of the nonzero submatrices here, 
\begin{align}
\text{sphere:} \hspace{20pt} \text{det}(\frac{\partial \chi_i}{\partial q_j})  &= \frac{1}{3^{3/2}}\\
\text{disk:} \hspace{20pt} \text{det}(\frac{\partial \chi_i}{\partial q_j}) &= \frac{\log(4)-1}{4}.
\end{align}

So the gauge-fixing factors should be in the two cases
\begin{align}
\frac{1}{\text{vol}(\psltwoc)}= \frac{1}{\text{vol}(\text{PSU}(2))}\frac{1}{3^{3/2}} \delta(\chi_1)\delta(\chi_2)\delta(\chi_3)\\
\frac{1}{\text{vol}(\psltwor)}= \frac{2^{3/2}}{\text{vol}(\text{U}(1))}\frac{\log(4)-1}{4}\delta(\chi_1)\delta(\chi_2).
\end{align}
We normalized the original transformations (\ref{transf}) so that with a unit measure, a full rotation has length $2\pi$. For the case of $U(1)$, this means simply $\text{vol}(\text{U}(1)) = 2\pi$. For the case of PSU(2), we can use the fact $\text{vol}(\text{PSU}(2)) = 2\pi \, \text{vol}(S^2) = 8\pi^2$.

\subsection{Putting the pieces together}
Putting the pieces together and dropping the divergent terms in the one-loop determinants (including the dangerous logarithmic piece, which will cancel between the two expressions), we find the following formulas
\begin{align}
\frac{Z_{\text{sphere}}}{\text{vol}(\psltwoc)} &= \left(\frac{\i e^{\frac{1}{b^2}}}{2\sin(\frac{\pi}{b^2})}(4\pi\mu)^{\frac{1}{b^2}+1}\right)\left( -\i \frac{\sqrt{\pi}\,bC}{\sqrt{2}}\cdot C^3 \cdot\frac{(\sqrt{\pi}b C)^{-\frac{11}{3}}}{e^{\frac{1}{2}\cdot 2.32713}}\right) \left(\frac{1}{8\pi^23^{3/2}}\right) \label{zsphereliouv} \\
\frac{Z_{\text{disk}}}{\text{vol}(\psltwor)} &= 
\left(\frac{-\i e^{\frac{1}{2b^2}(1 +2\pi \i)}}{e\sin(\frac{\pi}{b^2})}\left(\frac{\sqrt{4\pi\mu}}{\alpha}\right)^{\frac{1}{b^2}+1}\right)\left(  \frac{-\i\sqrt{\pi}\,bC}{\sqrt{1.51095}}\cdot C^2 \cdot\frac{(\sqrt{\pi}b C)^{-\frac{17}{6}}}{e^{\frac{1}{2}\cdot 1.14858}}\right) \left(\frac{2^{\frac{3}{2}}(\log(4)-1)}{2\pi\cdot 4}\right)\notag
\end{align}
We remind the reader that these formulas are valid in the semiclassical small $b$ limit, and further (for the disk) in the high-energy limit of small positive $\alpha$ (or large negative $\mu_B$). 
In these expressions, the first term is the classical action, the second term is the one-loop determinant, and the third term comes from the gauge fixing. The invariant ratio is
\be\label{ratio2}
\frac{Z_{\text{sphere}}}{\text{vol}(\psltwoc)} \cdot\left(\frac{\text{vol}(\psltwor)}{Z_{\text{disk}}}\right)^2 = 8.889\, e^{-\frac{2\pi \i}{b^2}} b \sin(\frac{\pi}{b^2})\, \alpha^{2 + \frac{2}{b^2}}.
\ee

Let's now apply this to the minimal string. To compute the partition functions of the minimal string, we set $b = \sqrt{2/p}$ where $p$ is an odd integer (and which must be large for our semiclassical approximation to be valid) and multiply by the partition function of the matter sector, which is the $(2,p)$ minimal model:
\be
\mathcal{Z}_{\text{sphere}} = Z_{\text{sphere}}^{\text{minimal model}}\frac{Z_{\text{sphere}}}{\text{vol}(\psltwoc)}, \hspace{20pt} \mathcal{Z}_{\text{disk}} = Z_{\text{disk $(1,1)$}}^{\text{minimal model}}\frac{Z_{\text{disk}}}{\text{vol}(\psltwor)}.
\ee
Using the formula (here $S_{(1,1),(1,1)}$ is an element of the modular S-matrix relating the identity characters in the two channels, see e.g.~\cite{DiFrancesco:1997nk},  Chapter 10)
\be
\frac{(Z_{\text{disk $(1,1)$}}^{\text{minimal model}})^2}{Z_{\text{sphere}}^{\text{minimal model}}} = S_{(1,1),(1,1)} = -\frac{2}{\sqrt{p}}\sin(\frac{\pi p}{2})\sin(\frac{2\pi}{p}),
\label{minimalmodelS11}
\ee
and approximating $\sin(\frac{2\pi}{p}) = \frac{2\pi}{p}$, we find that for large $p$ and small $\alpha$
\be\label{liouvillefinalanswer}
\frac{\mathcal{Z}_{\text{sphere}}}{(\mathcal{Z}_{\text{disk}})^2} = 1.000\, p \,\alpha^{p+2}.
\ee
We remind the reader that this formula is accurate in the limit of large $p$ and small $\alpha$. Here $p$ is an odd integer that labels the $(2,p)$ minimal string, and $\alpha$ is a parameter that determines the energy of the FZZT boundary condition. Small $\alpha$ corresponds to high energy. In the next section we will compute the same thing in the matrix integral language, and we will find that the numerical constant in (\ref{liouvillefinalanswer}) should be exactly one.

\section{Matrix integral computations}
A Hermitian matrix integral (see \cite{DiFrancesco:1993cyw}\cite{Eynard:2015aea} for reviews) is an integral of the form
\begin{align}
\mathfrak{Z} &= \int \d H \, e^{-L \ \text{Tr}V(H)},
\end{align}
where $L$ is the rank of the matrix and $V$ is the ``potential.'' This can be written as an integral over the eigenvalues,
\begin{align}\label{matrixEigs}
\mathfrak{Z} =C_L\int \mathrm{d}^L\lambda\, e^{-L\sum_{j = 1}^L V(\lambda_j)}\,\prod_{i<j}(\lambda_i - \lambda_j)^2.
\end{align}
Here the constant $C_L$ and the final term (Vandermonde determinant) both arise from integrating out the non-eigenvalue parts of the matrix. 

In the leading order at large $L$, one can formally ignore the discreteness of the eigenvalues and trade in the $L$ eigenvalues for a smooth density $\rho(\lambda)$, normalized so that $\int \d \lambda \rho(\lambda) = 1$:
\begin{align}
\mathfrak{Z} &\sim \int \mathcal{D}\rho  \, e^{-L^2 I[\rho]}\\
I[\rho] &= \int \d \lambda \rho(\lambda) V(\lambda) - \frac{1}{2}\int\hspace{-7pt}\int \d \lambda_1 \d \lambda_2 \log[(\lambda_1-\lambda_2)^2] \rho(\lambda_1)\rho(\lambda_2).
\label{iofrho}
\end{align}
In particular, for large $L$, we can think of the matrix integral as being dominated by a single saddle point $\rho_0(\lambda)$ which stationarizes this action, subject to the constraint $\int \d \lambda \rho_0(\lambda) = 1$. Because there are a total of $L$ eigenvalues, the total or ``physical'' density of eigenvalues is $L$ times this normalized density, so $L \rho_0(\lambda)$. This function is supported on an interval or a union of intervals, and generically it vanishes like a square root at the ends of each interval.

The $(2,p)$ minimal string theory is conjectured to be related to a type of matrix integral where $\rho_0$ is supported on the entire positive real axis, and with \cite{Moore:1991ir}
\be\label{rhoMS}
\rho_0(x) = \sinh\Big(\frac{p}{2}\text{arccosh}(1+2x)\Big).
\ee
This does not fit the definition of a standard matrix integral, because the density cannot be normalized so that its integral is one. However, it makes sense as an example of what is called a ``double scaled'' matrix integral. This can be defined as a limiting procedure applied to an ordinary matrix integral, where a family of potentials parametrized by $\epsilon$ are arranged so that very near the endpoint, the density locally approximates a rescaled version of (\ref{rhoMS}):
\be\label{confbackground}
\rho_0(\lambda) = \epsilon^{p/2}\sinh\Big(\frac{p}{2}\text{arccosh}(1 + 2E)\Big) + O(\epsilon^{p/2+1}), \hspace{20pt} \lambda = \lambda_{\text{endpoint}} + \epsilon E.
\ee
So in the limit $\epsilon \to 0$, we recover the full density (\ref{rhoMS}) in a ``zoomed-in'' view of a small neighborhood of one of the endpoints. ``Double scaling'' refers to following this limiting procedure, while also adjusting $L$ so that the total density of eigenvalues in the $E$ coordinate, which is proportional to 
\be
\epsilon^{p/2+1} L = e^{S_0},
\ee
is fixed. 
The result is a region near the edge of the spectrum that resembles (\ref{rhoMS}), attached to a larger ``garbage'' region at higher energies that depends on the details of the limiting procedure that was used.

What are we supposed to compute in this double scaled matrix integral? In the minimal string, we computed the sphere partition function, and we normalized it using the (FZZT) disk partition function in the high-energy limit. Both the sphere and disk  quantities have duals in the matrix integral picture. First, the sphere partition function is related to the leading term $L^2\mathcal{F}_0$ in the logarithm of the full matrix partition function:
\be
\log(\mathfrak{Z}) = L^2 \mathcal{F}_0 + \mathcal{F}_1 + L^{-2} \mathcal{F}_2 + \dots.
\ee
One can get this term by simply evaluating the action $I$ on the stationary configuration $\rho_0$:
\be\label{FREEEN}
\mathcal{F}_0 = -I[\rho_0].
\ee
Second, the (FZZT) disk partition function is given by a similar leading term $L\mathcal{G}_0$ in the expectation value
\be
\langle \text{Tr}\log(H-x)\rangle = L\mathcal{G}_0(x) + L^{-1} \mathcal{G}_1(x) + L^{-3} \mathcal{G}_2(x) + \dots
\ee
Again, this is given simply in terms of the stationary configuration $\rho_0$:
\be
\mathcal{G}_0(x) = \int \d\lambda \rho_0(\lambda) \log(\lambda-x).
\ee

The terms in the expansion that are proportional to negative powers of $L$ are well-defined in the double-scaled limit, in the sense that they do not depend on the ``garbage'' region that (\ref{rhoMS}) is attached to at high energies. However, the leading terms do depend on the garbage region. In fact, they are numerically dominated by it! There is a good analog of this in the Liouville path integral. The terms proportional to inverse powers of $L$ correspond to Liouville partition functions on surfaces with negative Euler characteristic. For such surfaces, the integral over the Liouville field converges along the real axis. But for the sphere or the disk (or, marginally, the torus), the integral is divergent in the large negative $\phi$ region. This corresponds to very small surfaces, and the ambiguity in how this part of the path integral is regulated corresponds to the ambiguity in the nonuniversal garbage that is used to construct the double-scaled limit of the matrix integral.

In Liouville, the nonuniversal pieces and the universal pieces can be distinguished by their dependence on the cosmological constant $\mu$. The contribution of the nonuniversal small $\sigma$ region is analytic in $\mu$, because for large negative $\sigma$, one can expand down in powers of $\mu e^{2\sigma}$, giving a power series in $\mu$.\footnote{The contour prescription we used for Liouville throws out these nonuniversal analytic parts automatically, but they would be there if for example we had defined the integral over $\sigma$ to be on a contour on the real axis that ended at some finite but large negative value, corresponding to a UV cutoff on the physical metric.} But as we saw above, the universal part depends on the $\mu e^{2\sigma}$ term in an essential way, and the result is proportional to a nontrivial power of $\mu$. So the interesting part of the answer can be selected by keeping the part that is nonanalytic in $\mu$.

In the matrix integral, $\epsilon^2$ plays the role of $\mu$, and the nonanalytic terms correspond to odd powers of $\epsilon$. As we will see, these terms are numerically highly subleading, but they are distinguished by their nonanalyticity as functions of $\epsilon^2$. For this to work, it is important that we do not accidentally introduce any direct nonanalyticity in $\epsilon^2$ through the matrix integral potential. So we will make sure that the potential is analytic in $\epsilon^2$, and we will then take the leading nonanalytic part of the free energy. Schematically, if the potential is chosen to be analytic in $\epsilon^2$, then we will have
\be
L^2 \mathcal{F}_0 = L^2(1 + \epsilon^2 + \epsilon^4 + \dots) + L^2 \epsilon^{p + 2}(1 + \epsilon^2 + \epsilon^4+\dots).
\ee
The leading nonanalytic piece is proportional to $L^2 \epsilon^{p+2} = e^{2S_0}$, and this piece can be identified as the universal part that can be compared to the universal part of the Liouville answer. There is a similar procedure for extracting the universal part of the FZZT partition function $\mathcal{G}_0(x)$, which we will describe below.

\subsection{The conformal background}
Let's now carry this out in detail. The first step will be to construct a family of potentials that gives the desired double-scaled limit. To do so it will be helpful to define a special set of functions that we will add together to get the desired density of states
\begin{align}
\rho_j(\lambda|a) = \frac{2^{2j}}{\binom{2j-1}{j-1}}\frac{(a^2-\lambda^2)^{j - \frac{1}{2}}}{2\pi a^{2j}}.
\end{align}
For each value of $j$, this is a normalized and symmetric density of states that extends between endpoints $\pm a$. Near one of these endpoints, the density of states behaves as
\be
\rho_j(\lambda|a) = \frac{2^{3j-\frac{1}{2}}}{\binom{2j-1}{j-1}}\frac{(\epsilon E)^{j-\frac{1}{2}}}{2\pi a^{j+\frac{1}{2}}} + \dots, \hspace{20pt} \lambda = -a+\epsilon E
\ee
where the dots are higher order in $\epsilon$. Because we can get different powers with different values of $j$, an appropriate linear combination of such functions with different coefficients can be used to approximate the conformal background near the edge.

The density of states $\rho_j(\lambda|a)$ is the large $L$ stationary configuration for a matrix integral with a particular potential $V_j(\lambda|a)$. Explicitly, this potential (or rather its first derivative) is 
\be
V'_j(\lambda|a) = (-1)^{j+1}\frac{2^{2j}}{\binom{2j-1}{j-1}}\frac{(\lambda^2 - a^2)^{j-\frac{1}{2}}_+}{a^{2j}}
\ee
where the subscript $(\cdot)_+$ means the terms proportional to non-negative powers of $\lambda$, when the expression is expanded around $\lambda = \infty$ (see,  for example,  section 2.2 of \cite{DiFrancesco:1993cyw}).
For example
\be
(\lambda^2-a^2)^{\frac{1}{2}}_+ = \lambda, \hspace{20pt} (\lambda^2 - a^2)^{\frac{3}{2}}_+ = \lambda^3 - \frac{3 a^2}{2}\lambda, \hspace{20pt} \dots
\ee
If we choose one of these potentials $V'_j(\lambda|a)$, then the resulting large $L$ density of states will be $\rho_j(\lambda|a)$. More generally, if we take a superposition with coefficients $c_i$ such that $\sum_{i}c_i = 1$, then the density of states will be the corresponding superposition:
\be\label{correspondence}
\rho_0(\lambda) = \sum_{j = 1}^\infty c_j \rho_j(\lambda|a) \hspace{20pt} \leftrightarrow  \hspace{20pt} V'(\lambda) = \sum_{j = 1}^\infty  c_jV_j'(\lambda|a).
\ee

In principle, we could use this to get the desired double scaled background (\ref{confbackground}), by setting $a$ to some value and then adjusting the coefficients $c_j$ as a function of $\epsilon$ so that near $\lambda = -a$, the density of states approximates (\ref{confbackground}). However, if we do this in the most straightforward way, the potential will not be analytic in $\epsilon^2$, and it will not be simple to isolate the desired term in the free energy.

To avoid this problem, we need to be more careful, adjusting the coefficients in the potential in a manifestly analytic way. As we will see, this will lead to an endpoint $a$ that is not analytic in $\epsilon^2$, contaminating the functions $V_k(\lambda|a)$ and making it difficult to impose analyticity. So rather than working with $V_j(\lambda|a)$, we will expand the potential in terms of $V_j(\lambda|1)$:
\be\label{potentialtk}
V'(\lambda) = \sum_{k = 1}^\infty t_k V'_k(\lambda|1).
\ee
This makes it easier to be sure that the potential is analytic: all we need to do is choose $t_k$ coefficients that are analytic in $\epsilon^2$. One can determine the density of states associated to this potential by the following procedure: (i) write $(\ref{potentialtk})$ in terms of $V'_j(\lambda|a)$ for an arbitrary $a$ to be fixed later, (ii) use (\ref{correspondence}) to get $\rho$, assuming that value of $a$ (iii) determine the correct value of $a$ by imposing the normalization constraint. For the first step, one can expand $V_k(\lambda|1)$ in terms of $V_j(\lambda|a)$ using
\be
(\lambda^2-1)^{k-\frac{1}{2}}_+ = \sum_{j = 0}^{k-1} \binom{k-\frac{1}{2}}{j}(a^2-1)^j (\lambda^2-a^2)^{k-j-\frac{1}{2}}_+
\ee
which follows from linearity of the $(\cdot)_+$ operator. Substituting into (\ref{potentialtk}) leads directly to 
\begin{align}\label{cjfromtk}
c_j &= a^{2j}\sum_{k = j}^\infty \binom{k}{j}(1-a^2)^{k-j}t_k.
\end{align}
These $c_j$ coefficients give the density of states, from (\ref{correspondence}). However, the result depends on an as-yet-undetermined endpoint $a$. This can be fixed by imposing the normalization condition
\be\label{normcond}
1 = \sum_{j = 1}^\infty c_j = \sum_{k = 1}^\infty \Big(1-(1-a^2)^{k}\Big)t_k.
\ee

Our goal is find a set of $t_k$ coefficients that are analytic in $\epsilon^2$, such that the procedure just outlined leads to a density of states that behaves as (\ref{confbackground}) near the endpoints. We will parametrize the odd integer $p$ in terms of an unrestricted integer $m$, so the relevant minimal model is
\be
(2,p) = (2,2m-1).
\ee
We claim that the solution for the $t_k$ parameters $t_1,\dots,t_m$ is\footnote{Up to a choice of convention, the terms that are explicitly written here are the nonzero KdV times of the conformal background \cite{Moore:1991ir}\cite{Turiaci:2020fjj}.}
\be\label{tksol}
t_k = \begin{cases} (-1)^{\frac{m-k}{2}}\frac{\binom{m-1}{\frac{m-k}{2}}\binom{k+m-2}{k}}{\binom{2m-2}{m}}\epsilon^{m-k}   + O(\epsilon^{m-k+2})& m-k \text{ even} \\ O(\epsilon^{m-k+1}) & m-k\text{ odd}\end{cases}
\ee
There is some freedom in choosing the subleading terms, but to get the correct density of states, we will have to choose them so that the solution to (\ref{normcond}) satisfies
\be
a^2 = 1-\epsilon + O(\epsilon^2).
\ee
For example, one valid choice is to set to zero the $O(\epsilon^{m-k+2})$ correction on the first line of (\ref{tksol}), and then to set $t_{k}$ for odd $k$ to be equal to minus $t_{k-1}$. Explicitly, this can be written as
\be
t_{m-2\ell} = -t_{m-2\ell+1} = (-1)^\ell\frac{\binom{m-1}{\ell}\binom{2m-2-2\ell}{m-2\ell}}{\binom{2m-2}{m}}\epsilon^{2\ell}.\label{texact}
\ee
A nice property of this particular choice is that the sum of the $t_k$ parameters telescopes, so that
\be\label{telescopes}
\sum_{k=0}^\infty t_k = 1
\ee
where we have formally defined
\be
t_0 = \begin{cases} (-1)^{\frac{m}{2}+1} \frac{\binom{m-1}{\frac{m}{2}}}{\binom{2m-2}{m}}\epsilon^{m} & m \text{ even} \\ 0 & m \text{ odd}\end{cases}.
\ee
This simplifies the normalization condition (\ref{normcond}) that determines $a$ to
\be
\sum_{k = 0}^\infty (1-a^2)^k t_k = 0,
\ee
which one can check is indeed solved by $1 - a^2 = \epsilon  + O(\epsilon^2)$.

It remains to check that near the endpoint, we get the desired density of states (\ref{confbackground}). For this one can neglect the higher order terms and use only the terms written explicitly in (\ref{tksol}). Using (\ref{cjfromtk}), we find
\begin{align}
c_j &= \frac{\binom{m}{j}\binom{m+j-2}{j-1}}{2^{j-m}\binom{2m-2}{m-1}}\epsilon^{m-j} +O(\epsilon^{m-j+1}).
\end{align}
This implies that near the endpoint at $a \approx -1$, we have
\begin{align}
\rho_0(\lambda)  &=\frac{2^{m+\frac{1}{2}}\epsilon^{m-\frac{1}{2}}}{2\pi} \sum_{j = 1}^m \frac{\binom{m}{j}\binom{m+j-2}{j-1}}{\binom{2j-1}{j-1}\binom{2m-2}{m-1}}(4E)^{j-\frac{1}{2}}  + O(\epsilon^{m+\frac{1}{2}})\\&= \frac{2^{m+\frac{1}{2}}\epsilon^{m-\frac{1}{2}}}{\pi\binom{2m-1}{m-1}}\sinh\left[\frac{2m-1}{2}\text{arccosh}(1+2E)\right] + O(\epsilon^{m+\frac{1}{2}}).\label{matrixDensity}
\end{align}
Up to an overall constant that can be absorbed into the definition of $e^{S_0}$, this is indeed the desired density of states.

\subsection{The free energy}
The explicit values (\ref{texact}) determine a potential exactly, which in turn determines $\rho_0$ exactly, and together these quantities determine the free energy $\mathcal{F}_0$ via for example (\ref{FREEEN}) or a somewhat more efficient version of this formula in \cite{Eynard:2015aea}. Using Mathematica, we found the first few cases:
\begin{align}
2\mathcal{F}_0^{(2,3)} &= -\log2 - \frac{25}{24} - \frac{\epsilon^2}{3} + \frac{\epsilon^4}{4} \boxed{- \frac{4\epsilon^5}{15}} + \dots \\
2\mathcal{F}_0^{(2,5)} &= -\log2 - \frac{49}{40}-\frac{\epsilon ^2}{5}+\frac{\epsilon ^4}{24}\boxed{-\frac{4\epsilon ^7}{105}} + \dots\\
2\mathcal{F}_0^{(2,7)} &= -\log2 - \frac{761}{560}-\frac{6 \epsilon ^2}{35}+\frac{8\epsilon ^4}{125}-\frac{\epsilon ^6}{50}+\frac{\epsilon ^8}{100}\boxed{-\frac{64\epsilon ^9}{7875}} + \dots\\
2\mathcal{F}_0^{(2,9)} &= -\log2 - \frac{7381}{5040}-\frac{10 \epsilon ^2}{63}+\frac{109 \epsilon ^4}{1372}-\frac{\epsilon ^6}{49}+\frac{3 \epsilon ^8}{392}\boxed{-\frac{64\epsilon ^{11}}{33957}} + \dots
\end{align}
Here we have boxed the first nonanalytic term in each case. This one term is the universal part of the answer and everything else is ``nonuniversal garbage,'' depending on specific decisions we made in constructing the double-scaled theory.

Note that in each case we have written $2\mathcal{F}_0^{(2,p)}$ on the LHS. This is because with the symmetric potential we have chosen, there is an identical copy of the same double-scaled theory at both edges of the spectrum. The matrix integral free energy includes contributions from both, but we intend $\mathcal{F}_0^{(2,p)}$ to mean the free energy associated to just a single copy, so $2\mathcal{F}_0^{(2,p)} = \mathcal{F}_0$.

It rapidly becomes impractical to calculate the answer this way, but in fact there is a simple general answer for arbitrary $p = 2m-1$:
\be\label{generalAns}
2\mathcal{F}_0^{(2,p)} = (\text{analytic in }\epsilon^2) -\frac{2^{p-1}p}{(p^2-4)\binom{p}{\frac{p-1}{2}}^2}\epsilon^{p+2} + O(\epsilon^{p+3}).
\ee
To derive this in an efficient way, one can use a formula, reviewed in appendix \ref{APPORTHOG}, that computes the free energy from the function $u(\xi) = 1-a^2(\xi)$ that solves the ``genus zero string equation''
\begin{align}
1-\xi &= \sum_{k = 0}^\infty t_k \Big(1-a^2\Big)^k\label{stringEQ}\\
&= \sum_{k = 0}^\infty t_k u^k\\
&\approx \frac{\epsilon^m}{2^m \binom{m-\frac{1}{2}}{m}}\Big(P_m(u/\epsilon) - P_{m-2}(u/\epsilon)\Big)\label{legendreP}\\
&=f(u).
\end{align}
In the second-to-last line, we used an approximate form for the $t_k$ that resums to a combination of Legendre polynomials, as pointed out in \cite{Turiaci:2020fjj}.
Concretely, this form neglects all of the unspecified higher-order terms in (\ref{tksol}), and it is a sufficiently good approximation to compute the leading nonanalytic term in the free energy. In the final line, we introduced a temporary notation $f(u)$. In solving this equation to get $u(\xi)$, we choose the branch of the solution that is equal to $\epsilon$ when $\xi = 1$. 

In terms of this function $u(\xi)$, the formula for the free energy is
\begin{align}
\mathcal{F}_0 &= \int_0^1 \d \xi (1-\xi)\log\left(\frac{1-u(\xi)}{4}\right)\label{ORTHOG}\\
&=\int_1^\epsilon (-\d u f'(u)) f(u) \log(\frac{1-u}{4}).
\end{align}
The function $f(u)$ is analytic in $\epsilon^2$, so for generic $\xi$, the solution $u(\xi)$ will also be analytic in $\epsilon^2$. However, this breaks down near $\xi = 1$, where the solution is $u = \epsilon$. So, to accurately compute the nonanalytic terms, we only need to do the integral in the vicinity of $\xi = 1$, which corresponds to $u$ in the vicinity of $\epsilon$, where $\log(1-u)\approx -u$. Using $\sim$ to denote equality of the leading nonanalytic terms, we therefore have
\begin{align}
\mathcal{F}_0& \sim  \int_?^\epsilon (-\d u f'(u))f(u)(-u)\\
&\sim \int_0^\epsilon (-\d u f'(u))f(u)(-u).
\end{align}
In the first line, we introduced an arbitrary lower limit of integration. As long as this point is chosen to be analytic in $\epsilon^2$, it will not affect the nonanalytic terms in the answer, and in the second line we chose a convenient value of zero. As we will see, this has the nice effect of removing completely the analytic terms, leaving only the nonanalytic part that we are seeking. After integrating by parts, we continue
\begin{align}
 &= -\int_0^\epsilon \d u \frac{f^2(u)}{2}\\
&=-\frac{\epsilon}{2}\left(\frac{\epsilon^m}{2^m\binom{m-\frac{1}{2}}{m}}\right)^2\left(\frac{1}{2m+1} + \frac{1}{2m-3}\right).
\end{align}
To get the final line, we inserted the expression for $f(u)$ in (\ref{legendreP}) and used 
\be
\int_0^1 \d y P_m(y)P_m(y) = \frac{1}{2m+1}, \hspace{20pt} \int_0^1 \d y P_m(y)P_{m-2}(y) = 0
\ee
which follow from the orthogonality relation for the $P_m$ functions, and the fact that $P_m$ and $P_{m-2}$ are either both even functions or both odd functions. After substituting in $p = 2m-1$, one finds the term in (\ref{generalAns}).

\subsection{FZZT disk}
We start by reminding the reader of the definition
\be
\mathcal{G}_0(x) = \int_{-a}^a \d \lambda \rho_0(\lambda)\log(\lambda - x).
\ee
To define a function appropriate for the double-scaled limit, we consider this function at an argument $x$ that is close to the lower endpoint. For example, in the case of the $(2,3)$ model, we have the explicit formula (after giving $E$ a small negative imaginary part)
\be\label{disk23}
\mathcal{G}_0^{(2,3)}(-a + \epsilon E) = \frac{7}{12}-\frac{2(1+2E)}{3}\epsilon +\left(\frac{2}{3}+\frac{(1+2E)^2}{2}\right) \epsilon ^2+\boxed{\i\frac{2^{7/2}}{15} E^{3/2} (5+4 E) \epsilon^{5/2}}+O(\epsilon^3).
\ee
Here we have boxed the universal term, which will be compared to the Liouville computations below. We expect that everything else in this expression is ``nonuniversal garbage'' which depends on the way in which we take the double-scaled limit.

At first, this might seem puzzling, because the ``nonuniversal garbage'' contains a term $\epsilon^1$, which would seem to be a nonanalytic function of $\mu \sim \epsilon^2$. However, from the Liouville perspective, precisely the combination $(1+2E)\epsilon$ is proportional to $\mu_B$, the boundary cosmological constant, and we should expect nonuniversal analytic terms in both $\mu$ and $\mu_B$, associated to the divergence of the path integral in the large negative $\sigma$ region.\footnote{Again, the contour we used for Liouville effectively set all such terms to zero, but with a different contour prescription, such terms would be present.} In Liouville language, the term at order $\epsilon^2$ is a linear combination of $\mu$ and $\mu_B^2$. By contrast, the boxed term is genuinely nonanalytic in $\mu_B$ and $\mu$, and corresponds to a contribution from the universal continuum region of the Liouville path integral.

In the Liouville computation, the disk path integral was pure imaginary, which suggests that the first nonanalytic term (in this sense) will be pure imaginary. This is true of (\ref{disk23}), and also true for the $(2,5)$ and $(2,7)$ cases, which we checked explicitly. We don't have a general proof of this from the matrix side, although we suspect it is possible to show this. However, what we can do easily is compute the imaginary part:
\begin{align}
\text{Im}\,\mathcal{G}_0(-a+\epsilon E) &= \text{Im} \int_{-a}^a \d \lambda \rho_0(\lambda)\log(\lambda+a-\epsilon E) \\
&= \pi \int_{-a}^{-a+\epsilon E} \d\lambda \rho_0(\lambda).
\end{align}
This only depends on the density of states near the edge, where it is constrained by the double-scaled limit. Putting in (\ref{matrixDensity}) and taking high energies $E \gg 1$, this is
\begin{align}\label{diskmatrixanswer}
\text{Im}\,\mathcal{G}_0(-a+\epsilon E) \to \frac{(2\epsilon)^{\frac{p}{2}+1}}{4\binom{p}{\frac{p-1}{2}}}\frac{(4E)^{\frac{p}{2}+1}}{p+2}.
\end{align}

\subsection{Comparing the ratio to the Liouville answer}
As a final step, we need to relate the $\alpha$ parameter of the Liouville theory to the energy $E$. We can compare $Z_{\text{disk}}'(\mu_B)$ with $\mathcal{G}_0'(x)$, which is proportional to $\rho_0(E)$. For this we need to study the theory at finite energy, where we did not compute the one-loop determinant. Fortunately, the classical answer will be enough. In the classical approximation (which means only keeping terms of order $\frac{1}{b^2} = \frac{p}{2}$ in the exponential), the $\alpha$ dependence of $Z_{\text{disk}}'(\mu_B)$ is proportional to
\be
e^{-\frac{1}{b^2}\log(\alpha)} = e^{\frac{p}{2}\log(\frac{1}{\alpha})}.
\ee
On the other hand, for large $p$, the density $\rho_0(E)$ is proportional to
\be
e^{\frac{p}{2}\text{arccosh}(1+2E)}.
\ee
Comparing the two, we conclude that\footnote{With (\ref{hemisphereSol}), this implies $
-\sqrt{\frac{\pi}{\mu}}\mu_B = 1 + 2E$,
which justifies a statement made above that $(1+2E)\epsilon \propto \mu_B$.}
\be
\label{equivto}
\log\frac{1}{\alpha} = \text{arccosh}(1+2E). 
\ee
Or, at high energies,
\be
\frac{1}{\alpha} \approx 4E.
\ee
Substituting this into (\ref{diskmatrixanswer}), and also using (\ref{generalAns}), we find
\be
\frac{\mathcal{F}_0^{(2,p)}|_{\text{universal}}}{(\mathcal{G}_0^{(2,p)}|_{\text{universal}})^2} = \frac{p+2}{p-2}p\, \alpha^{p+2}, \hspace{20pt} \alpha \ll 1.
\ee
This exact answer agrees with (\ref{liouvillefinalanswer}) at large $p$.

\section*{Acknowledgements} 
RM is supported in part by Simons Investigator Award \#620869.
DS was supported in part by DOE grant DE-SC0021085. 

\appendix

\section{The sphere partition function in JT gravity is infinite}
In the main text, we found that the matrix integral density
\be\label{mainTextanswer}
\rho^{\text{total}}(E)\d E = L\frac{(2\epsilon)^{\frac{p}{2}+1}}{\pi\binom{p}{\frac{p-1}{2}}}\sinh\left[\frac{p}{2}\text{arccosh}(1+2E)\right]\d E
\ee
corresponds to the leading term in the free energy
\be\label{mainTextFree}
\log(\mathfrak{Z}) \supset -L^2\frac{2^{p-2}p}{(p^2-4)\binom{p}{\frac{p-1}{2}}^2}\epsilon^{p+2}.
\ee
The density that approximates JT gravity \cite{Teitelboim:1983ux}\cite{Jackiw:1984je}\cite{almheiri2015models}\cite{Jensen:2016pah}\cite{Maldacena:2016upp}\cite{engelsoy2016investigation} in the large $p$ limit is \cite{Saad:2019lba}
\be
\frac{e^{S_0}}{2\pi^2}\sinh\left[\frac{p}{2}\text{arccosh}\left(1 + \frac{8\pi^2}{p^2}E_{\text{JT}}\right)\right] \d E_{\text{JT}} \approx \frac{e^{S_0}}{2\pi^2}\sinh\left(2\pi \sqrt{E_{\text{JT}}}\right) \d E_{\text{JT}}.
\ee
This can be obtained from (\ref{mainTextanswer}) by setting
\be
L\frac{(2\epsilon)^{\frac{p}{2}+1}}{\pi\binom{p}{\frac{p-1}{2}}}\d E = \frac{e^{S_0}}{2\pi^2}\d E_{\text{JT}} = \frac{e^{S_0}}{2\pi^2}\frac{p^2}{4\pi^2}\d E.
\ee
Substituting this into (\ref{mainTextFree}), we find
\be
\log(\mathfrak{Z}) \supset -\frac{e^{2S_0}}{2^{10}\pi^6}\frac{p^5}{p^2-4}.
\ee
This diverges in the large $p$ limit where the $(2,p)$ minimal string becomes JT gravity. This implies that the JT gravity sphere partition function is infinite, as suggested in \cite{Maldacena:2019cbz}.

\section{The free energy from orthogonal polynomials}\label{APPORTHOG}
The orthogonal polynomials for a given potential $V$ are defined so that the leading term in each polynomial is $p_n(\lambda) = \lambda^n + \dots$, and so that they are orthogonal to each other:
\be
\int_{-\infty}^\infty \d \lambda e^{-V(\lambda)}p_n(\lambda)p_m(\lambda) = s_n \delta_{nm}.
\ee 
Since the normalization is fixed by saying that the coefficient of the $\lambda^n$ term is one, the normalization $s_n$ in this equation is meaningful.

There is a simple formula for the free energy of the matrix integral in terms of this data, as reviewed in section 2.3 of \cite{DiFrancesco:1993cyw}. The leading $L^2$ term in the free energy is
\be\label{FREEENORTHOG}
\mathcal{F}_0 = \int_0^1 \d \xi (1-\xi)\log f(\xi)
\ee
where $\xi = n/L$ is a continuum version of the index $n$ of the orthogonal polynomials, and where $r(\xi) = s_n/s_{n-1}$. As reviewed in section 2.4 of \cite{DiFrancesco:1993cyw}, the function $r(\xi)$ satisfies an equation
\be
\xi = \sum_{m\text{ odd}}v'_m \binom{m}{\frac{m+1}{2}}r^{\frac{m+1}{2}}
\ee
where the derivative of the potential is parametrized as $V' = \sum_{m\text{ odd}}v'_m \lambda^m$. For a potential of the form (\ref{potentialtk}), this equation is
\be
\xi = \sum_{k = 1}^\infty t_k (1 - (1-4r)^k).
\ee
Writing $4r = a^2$, and using (\ref{telescopes}) this is the equation quoted in (\ref{stringEQ}), and the formula (\ref{FREEENORTHOG}) above becomes the formula (\ref{ORTHOG}) in the main text.

\bibliography{references}

\bibliographystyle{utphys}

\end{document}